\documentclass[onecolumn,superscriptaddress,secnumarabic,amssymb,nobibnotes,aps]{revtex4-1}
\usepackage{amssymb,amsmath,graphicx,amsthm,wrapfig,caption,epstopdf,afterpage}

\newcommand{\eps}{\epsilon}
\newcommand{\bra}[1]{\left(#1\right)}
\newcommand{\Bra}[1]{\left[#1\right]}
\newcommand{\BRA}[1]{\left\{#1\right\}}

\newcommand{\parf}[2]{\frac{\partial #1}{\partial #2}}

\begin{document}

\title{Pattern formation aspects of electrically charged tri-stable media with implications to bulk heterojunction in organic photovoltaics}

\author{Alon Z. Shapira}
\affiliation{Department of Solar Energy and Environmental Physics, Swiss Institute for Dryland Environmental and Energy Research, Blaustein Institutes for Desert Research, Ben-Gurion University of the Negev, Sede Boqer Campus, Midreshet Ben-Gurion 8499000, Israel}
\author{Nir Gavish}
\affiliation{Department of Mathematics, Technion - IIT, Haifa, 3200003, Israel}
\author{Arik Yochelis}
\affiliation{Department of Solar Energy and Environmental Physics, Swiss Institute for Dryland Environmental and Energy Research, Blaustein Institutes for Desert Research, Ben-Gurion University of the Negev, Sede Boqer Campus, Midreshet Ben-Gurion 8499000, Israel}
\affiliation{Department of Physics, Ben-Gurion University of the Negev, Be'er Sheva 8410501, Israel}

\begin{abstract}
	A common thread in designing electrochemically-based renewable energy devices comprises materials that exploit nano-scale morphologies, e.g., supercapacitors, batteries, fuel cells, and bulk heterojunction organic photovoltaics. In these devices, however, Coulomb forces often influence the fine nano-details of the morphological structure of active layers leading to a notorious decrease in performance. By focusing on bulk heterojunction organic photovoltaics as a case model, a self-consistent mean-field framework that combines binary (bi-stable) and ternary (tri-stable) morphologies with electrokinetics is presented and analyzed, i.e., undertaking the coupling between the spatiotemporal evolution of the material and charge dynamics along with charge transfer at the device electrodes. Particularly, it is shown that tri-stable composition may stabilize stripe morphology that is ideal bulk heterojuction. Moreover, since the results rely on generic principles they are expected to be applicable to broad range of electrically charged amphiphilic-type mixtures, such as emulsions, polyelectrolytes, and ionic liquids.
\end{abstract}

\maketitle

	\section{Introduction} Self-organization is a universal property in which spatial and/or temporal order/correlation in systems (or certain symmetry) emerges from interaction of individual subsets at much lower scales~\cite{WhGR:2002}. Unlike autocatalytic process that often give rise to complex temporal dynamics in biological, ecological and chemical systems~\cite{Cross1993851,Pismen2006,Meron2015}, applications related to energy and material science view the emerging self-assembled phenomena as gradient flows~\cite{nepomnyashchy2015coarsening}, i.e., systems that evolve via overall free energy minimization~\cite{Emmerich2012665}. For the latter, the two fundamental ingredients incorporate the thermodynamic free energy and the potential energy dissipation, here it is convenient to imagine a binary system of two immiscible phases, e.g., oil and water or di-block copolymers, that exhibits asymptotic phase separation. In case the phases are additionally electrically charged, e.g., electrochemical applications, the bulk evolution is driven by competing short-range and long-range interactions and results in rich diversity of periodic lamellar- and hexagonal-like morphologies~\cite{ohta1986equilibrium,kawasaki1988equilibrium,choksi2005diblock}. Moreover, recently it was found that these systems exhibit also several universal properties of dissipative media~\cite{gavish2016theory,bier2017bulk,gavish2017solvent}, such as formation of coexisting localized states via homoclinic snaking~\cite{gavish2017spatially}, which not only challenges our understanding of electrically charged self-assembly but also has implications to material science and applications thereof. 
	
	The ability to design and control over a broad range of meso- and nano-scale morphologies in organic materials make them attractive to many applications and specifically to renewable energy applications, examples of which include incorporation of ionic liquids in super-capacitors and batteries~\cite{Armand2009621,Wishart2009956,hayes2015structure}, polyelectrolyte membranes used in fuel cells~\cite{Morkved1996931,nyrkova1994microdomain,mauritz2004state}, and photovoltaics (OPV)~\cite{ChFi:2010,OPV_Rev:2011,Dang20133734}. The latter are particularly intriguing since unlike inorganic solar cells, they are involved with creation and annihilation of an electrically neutral particle -- an exciton -- that is generated upon illumination~\cite{GrHa:2003,HQL:2011}. Excitons are short-lived charge pairs with a typical diffusion length of roughly 10 nm.  {Therefore, OPV efficiency critically relies on an intertwined (mosaic) morphology of electron donor and acceptor materials of about exciton diffusion length, enabling the majority of the generated excitons to reach the donor/acceptor interface and dissociate to free charges that are then collected at the electrodes. This fine nano-structure, a.k.a. bulk heterojunction (BHJ), increases the efficiency of the OPV to a level competitive with that of inorganic devices~\cite{HQL:2011,C2JM33645F,GKAB:2012,CTA:2011,POLB:POLB23063,KVOWHG:2011,PhysRevLett.108.026601}}.
	
	Performance and efficiency of an BHJ OPV depends among many factors also on integrity of the labyrinthine-type morphology~\cite{CGGHMA:2010,ZSvAMVvM:2009,PhysRevLett.108.026601,AENM:AENM201000023,kouijzer2013predicting,treat2014phase,cardinaletti2014toward,vongsaysy2014formulation,zhou2015phase,mateker2017progress}. {Indeed, even 1nm  changes in the acceptor or donor labyrinth width can lead to a significant decrease in overall efficiency.  Nevertheless, following the assumption} that morphology evolution is much slower than other degradation mechanisms (e.g., oxidation), {existing OPV models consider `frozen' morphologies}~\cite{Buxton2006,Ray2012204,RCL:2013,groves2016simulating}. Yet, since exciton dissociation is responsible for electrical charge and, thus, for electrical force, the interrelation between interface properties and charge separation kinetics may become equally important to BHJ evolution, including instability. Furthermore, recent experimental evidences report on identification of an intermediate (third) phase in between the donor and the acceptor~\cite{Reid201227,razzell2013directly,Bartelt2013364,Burke20141923,muller2014active,gasparini2016designing}, which is often also being referred to as ``spaghetti''or ``river and streams'' phase~\cite{yin2011new}. The average donor and acceptor concentrations throughout the ``spaghetti'' phase are about the same. Yet, this intermediate phase is distinct from a randomly distributed amorphous composition, as it posses correlations of perculative nature and is, therefore, regarded as a third phase of the material. It is believed that properties of this third phase, e.g., composition and width, are strongly related to efficiency of exciton dissociation and also implicitly to the charge transport toward the electrodes~\cite{CTA:2011,yin2011new,mukherjee2015significance}. As such, decoupling between Coulombic forces and the interfacial BHJ evolutions cannot be neglected~\cite{Buxton2006}.
	
	Here we extend a continuum (mean field) framework introduced Buxton-Clarke~\cite{Buxton2006} by incorporating morphology evolution and accounting for the possible emergence of a stable intermediate phase in between the pure donor and acceptor phases. The outcome is a self-consistent model which unifies morphology and electrokinetics by incorporating and coupling three components: multi interface development employing Onsagers' approach to phase separation, charge transport via modification of Poisson-Nernst-Planck framework, and bimolecular interaction terms for exciton dynamics. Through bifurcation analysis we obtain a parameter plane for coexisting front and periodic solutions involving both two-phases and three-phases, and show that the width of the intermediate phase is in fact independent on domain size. {The analysis is then applied to OPV  efficiency in bilayer and ideal stripe morphologies}. The approach not only provides a systematic road map to understanding BHJ in OPV but also insights to degradation mechanisms related to morphological instabilities and a methodology to tackle three-phase compositions in material science applications. 
	
	\section{Model equations}
	We start with by formulating model equations in which the evolution of donor/acceptor morphology depends on the charge charier concentrations and vise verse. The donor/acceptor morphology is described by an order parameter $u:=\varphi_A-\varphi_D$, where $\varphi_A$ and $\varphi_D$ are volume fractions of acceptor and donor, respectively, so that $u$ varies from $u\equiv u_- = -1$ for {\it pure donor} to $u\equiv u_+ = 1$ for {\it pure acceptor}. The electrical charges are holes ($p$) and electrons ($n$), while in addition there are also neutral excitons ($\chi$) from which the charges are dissociated. Finally, the system is complemented by Poisson's equation. The model equations are derived from a free energy (see SI for details) and also comprise exciton dissociation/decay kinetics as well as electron-hole recombination and thus keep fidelity to the Buxton-Clarke framework~\cite{Buxton2006}. Due to interest in phenomenological properties of donor/acceptor interfaces, we present here the equations in their dimensionless form and refer the reader to SI for details:
\begin{widetext}
	\begin{subequations} \label{eq:rmodel}
		\begin{eqnarray}
		\parf{u}{t}&=&\underbrace{D_u\nabla\cdot\BRA{{\nabla u+}\bra{1-u^2} \Bra{\bra{\beta W''(u)\nabla u-\lambda\nabla^3u}}}}_{\rm phase~separation} 
			 +\underbrace{D_u \zeta\nabla\cdot \BRA{ \bra{1-u^2} \Bra{\bra{p+n}\nabla u+\bra{1+u}\nabla p-\bra{1-u}\nabla n}}}_{\rm donor/acceptor~affinity~to~charges}
			\label{eq:ru}, \\
			\parf{\chi}{t}&=&\underbrace{\nabla^2 \chi}_{\rm diffusion}-\underbrace{\frac{1-u^2}{\tau}\chi}_{\rm dissociation}-\underbrace{\chi}_{\rm decay}\label{eq:rchi}+\underbrace{G}_{\rm generation}, \\
			\parf{p}{t}&=&D_p\nabla\cdot\Bra{\underbrace{p\nabla\phi+\nabla p}_{\rm drift-diffusion}+\underbrace{\zeta p(1+u)\nabla u}_{\rm charge~ affinity}}+\frac{1-u^2}{\tau}\chi-\underbrace{\gamma\,np}_{\rm recombination}, \label{eq:rp} \\
			\parf{n}{t}&=&D_n\nabla\cdot\Bra{-n\nabla\phi+\nabla n-\zeta n(1-u)\nabla u}+\frac{1-u^2}{\tau}\chi-\gamma\,np, \label{eq:rn}\\
			&&\nabla \cdot \Bra{\eps \nabla \phi}=n-p\label{eq:rphi},
		\end{eqnarray}
		\end{subequations}
\end{widetext}
	where $W(u)=(1-u^2)^2(u^2+\xi/2)/2$ and other details are given in the SI. 
	
	The first term in~\eqref{eq:ru} is a typical Cahn-Hilliard approach to phase separation (in the absence of charges) but here we introduce a transition from double- to triple-well potential when $\xi<\xi_0=(1+\beta)/\beta$, so that in addition to pure donor/acceptor phases the system stabilizes also at $u\equiv u_0$ (here $u_0=0$) that corresponds to an intermediate phase; in what follows we refer to this phase as the \textit{intermediate phase}. 
	{Exciton dissociation rate is phase dependent and is maximized at the intermediate phase}, see \eqref{eq:rchi}, {so that the length scale of the intermediate phase is also affected by the second term in~\eqref{eq:ru} which represents coupling to electrons and holes, similarly to interaction in the context of charged polymers~\cite{tsori2007phase}}. The other equations~\eqref{eq:rp} and ~\eqref{eq:rn} comprise standard terms with additional contribution to transport of charge preference to respective phases, i.e., charge affinity. Model~\eqref{eq:rmodel} not only capture all the qualitative features of charge generation-recombination and transport that have been suggested by Buxton and Clarke~\cite{Buxton2006} but also allow coupling between Coulombic forces to morphology evolution.  Notably the {introduction of the intermediate phase does not require using} Onsager's phenomenological theory for dissociation and Poole-Frenkel mobility, that have been used to evaluate charge dynamics over a frozen double-well media~\cite{Buxton2006}. This distinct approach provides a framework that is not limited to specific functional forms, the formulation is flexible and allows physical refinement by demand. The qualitative features of one- and two-space dimensional solutions, however, will be shown as generic since they rely on domain walls (or what is mathematically being referred to as heteroclinic connections), so that the quantitative regimes can change while the qualitative properties persist.
	
	\section{Analysis}
	Equations~\eqref{eq:rmodel} have multiplicity of uniform solutions due to conservation of the order parameter $u$, see~\eqref{eq:ru}, which are given by:
	\begin{equation}
	\left(\begin{array}{c}
	u_0\\\chi_0\\p_0\\n_0\\\phi_0
	\end{array}\right)=\left(\begin{array}{c}
	u\\\frac{{\tau G}}{\tau+1-u^2}\\\sqrt{\frac{G}{\gamma}\frac{1-u^2}{\tau+1-u^2}}\\\sqrt{\frac{G}{\gamma}\frac{1-u^2}{\tau+1-u^2}}\\0
	\end{array}\right),
	\end{equation}
	and determined by selection of $u$. As for the typical bi-stable system (double well potential, $\xi>\xi_0$) also here the pure phases $u= u_\pm \lesssim 1$ are linearly stable while $u_0$ is linearly unstable for any $G>0$. Stability properties of $u_0$ for the tri-stable case, i.e., for $\xi<\xi_0$, are different as will be shown next. 
	
\begin{figure*}[tp]
		(a)\includegraphics[width=0.465\textwidth]{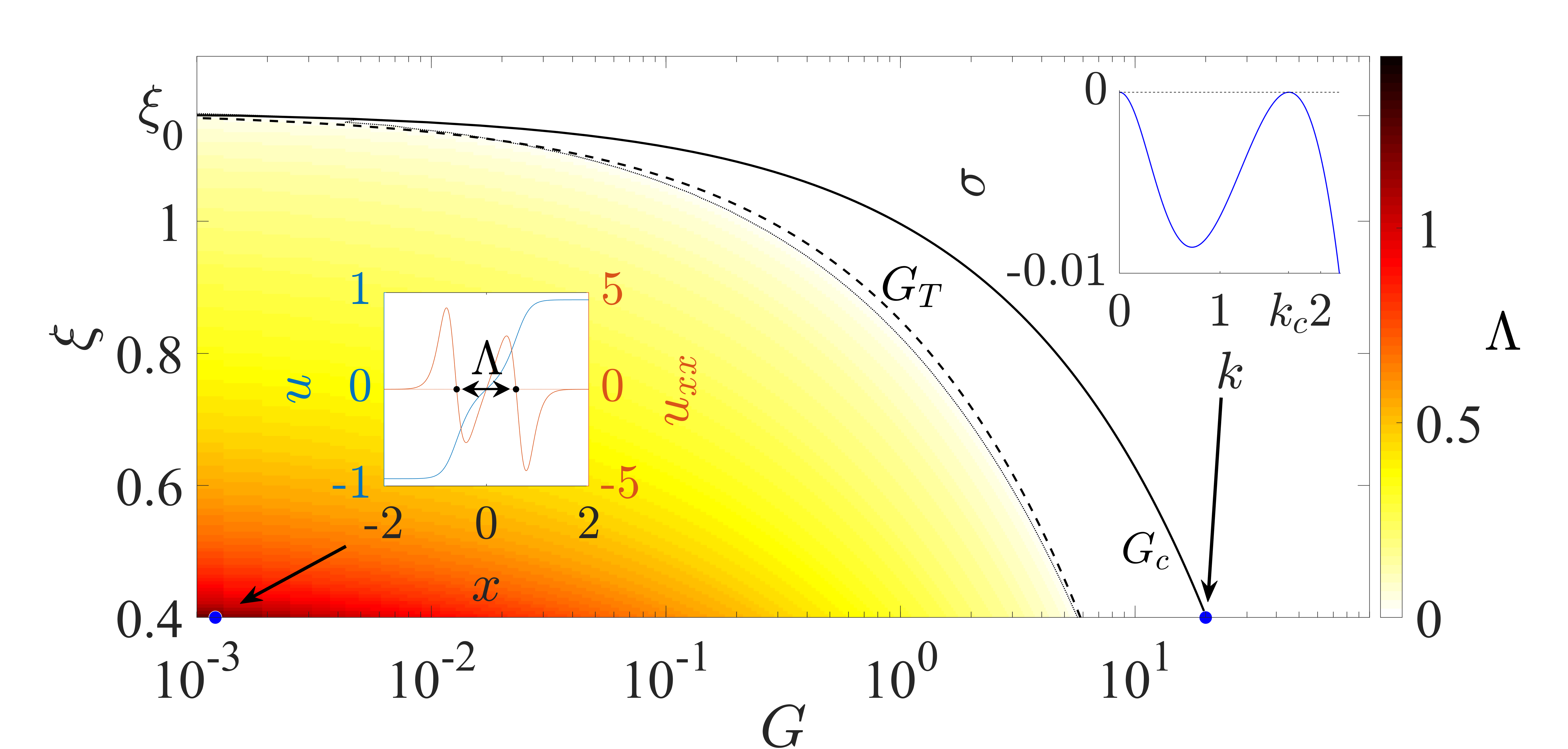} 
		(b)\includegraphics[width=0.465\textwidth]{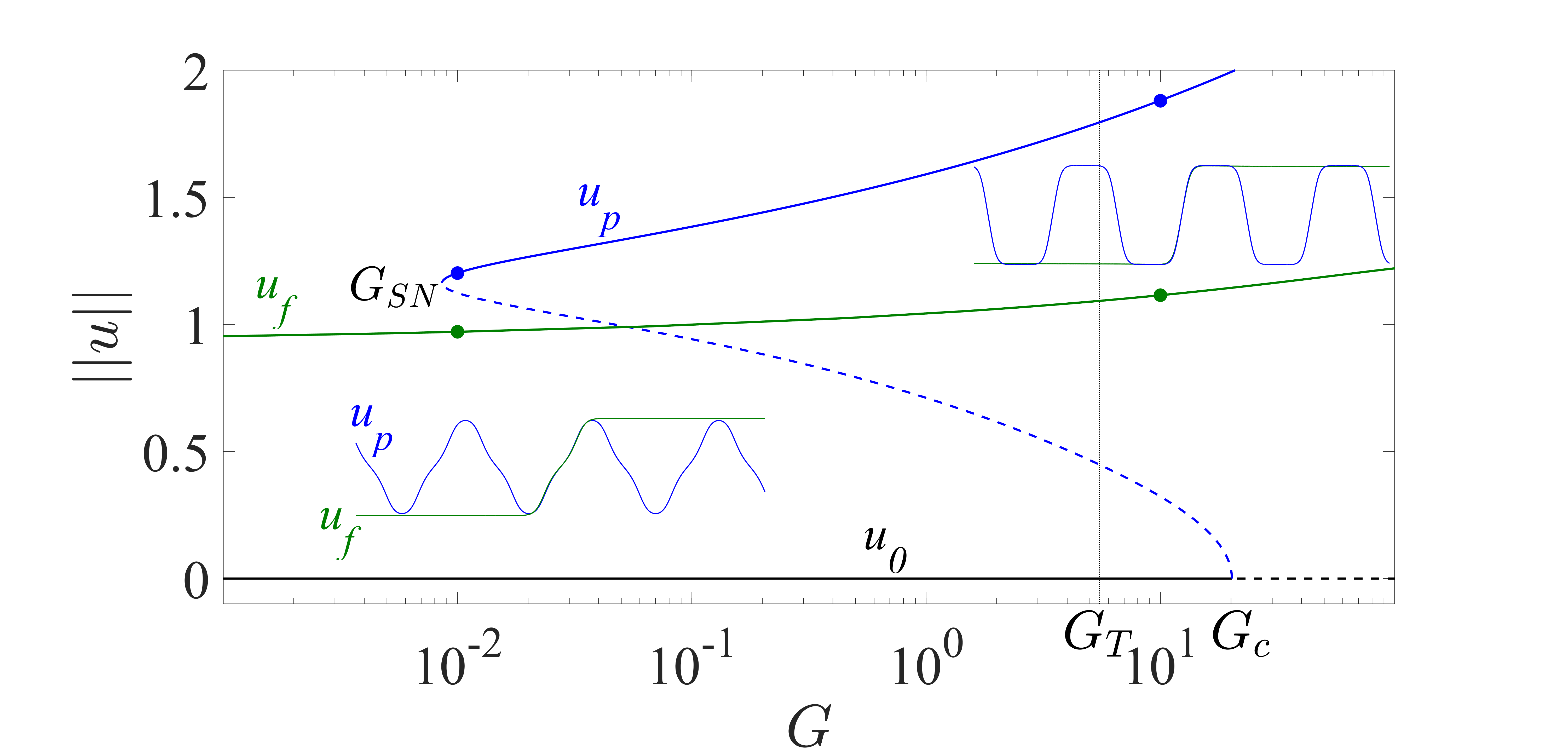}
		\caption{(a) Parameter space showing the onset of a finite wavenumber instability (solid line) and region where three-phase solutions exist $G<G_T$ (shaded area), where the dotted line obtained by numerical integration and dashed line according to~\eqref{gT}. The left inset shows the width of the intermediate phase, $\Lambda$, by solving for zeros of $u''=0$ with $x\neq 0$ (see the two limiting dots) while the right inset depicts the dispersion relation as the onset $G=G_c$ according to \eqref{gc}. (b) Representative bifurcation diagram across dots in (a), i.e., for {$\xi=0.4$}, showing the coexisting front ($u_f$) and sub-critically emerging, from $u=u_0$, periodic ($u_p$) solutions as a function of exciton generation ($G$), where {$||u||\equiv \sqrt{L^{-1}\int{[u^2+(u')^2]}\,{\rm d}x}$}, $L$ being the domain size (in the case of $u_p$ it is one period), and $G_{SN}$ is the saddle-node bifurcation; solid lines indicate linear stability. Solution branches have been obtained using numerical continuation via the \emph{pde2path} package~\cite{uecker2014pde2path,dohnalpde2path}, using periodic and no-flox (Neumann) boundary conditions for $u_p$ and $u_f$, respectively. The insets show three- and two-phase superimposed front and periodic solutions at locations indicated by dots, respectively; the plotted ranges are $u\in[-1,1]$ (vertical axis) and $x\in[0,12]$ (horizontal axis). The dotted line marks the transition from three- to two-phase solutions. Parameters: $\lambda=0.3$, $\beta=6$, $\zeta=9$, $\tau=100$, $\gamma=100$, $\epsilon=0.2$, $D_u=0.01$, $D_p=2$, $D_n=1$.} \label{fig:fig1}
	\end{figure*}

	To keep fidelity to OPV, we use $G>0$ (exciton generation that is related to illumination) and $\xi$ (depth of the intermediate well) as control parameters, as shown in Fig.~\ref{fig:fig1}(a). While linear stability of pure phases $u=u_\pm$ persists everywhere and thus existence of front solutions, i.e., solutions for which $u \to u_\pm$ as $x\to \pm \infty$, linear stability analysis about $u_0$ in one-space dimensions of the form~\cite{gavish2016theory,bier2017bulk,gavish2017spatially}
	\begin{equation}\label{pert}
	\left(\begin{array}{c}
	u\\\chi\\p\\n
	\end{array}\right)-\left(\begin{array}{c}
	0\\\chi_0\\p_0\\n_0
	\end{array}\right)\propto \text{e}^{\sigma t+\text{i}kx}+{\rm complex~conjugated},
	\end{equation}
	and substitution of the linearized Poisson's equation into other fields, shows that one dispersion relation corresponds to a finite wavenumber in stability, namely at $G=G_c$ the growth rate $\sigma$ of the perturbations is real and satisfies $\sigma(0)=\sigma(k_c)=0$ and otherwise $\sigma(k)<0$, for all $k$.
	This instability is, however, coupled to the marginal Goldstone mode $k=0$ due to conservation of the order parameter, i.e., $\sigma(0)=0$ always persists. The right inset in Fig.~\ref{fig:fig1}(a) shows a typical dispersion relation for $G>0$ and $\xi<\xi_0$.
	
	While the dispersion relations are cumbersome and obtained numerically, the critical values at the onset can be derived:
	\begin{equation}\label{gc}
	G_c=\frac{\epsilon^2\gamma(\tau+1)(\beta\xi-\beta-1)^2}{{4(\lambda-\epsilon\zeta+\epsilon\zeta^2-2\zeta\sqrt{\epsilon\lambda})^2}},
	\end{equation}
	\begin{equation}\label{kc}
	k^2_c=\frac{(1-\zeta\sqrt{\epsilon/\lambda})(\beta\xi-\beta-1)}{{\lambda-\epsilon\zeta+\epsilon\zeta^2-2\zeta\sqrt{\epsilon\lambda}}}.
	\end{equation}
	Consequently, the onset properties within this model are found to be independent of diffusion coefficient. On the other hand, while the onset $G_c$ depends linearly on the recombination rate ($\gamma$) and the dissociation rate ($\tau$) there is no dependence on the periodicity, i.e., change in $k_c$. The latter indicates that excitons play a negligible role in periodicity of the bifurcating periodic solutions, $u_p$, which are determined mostly by material properties and interaction between the material and charges, i.e., affinity.
	
\begin{figure*}[tp]
		\includegraphics[width=0.7\textwidth]{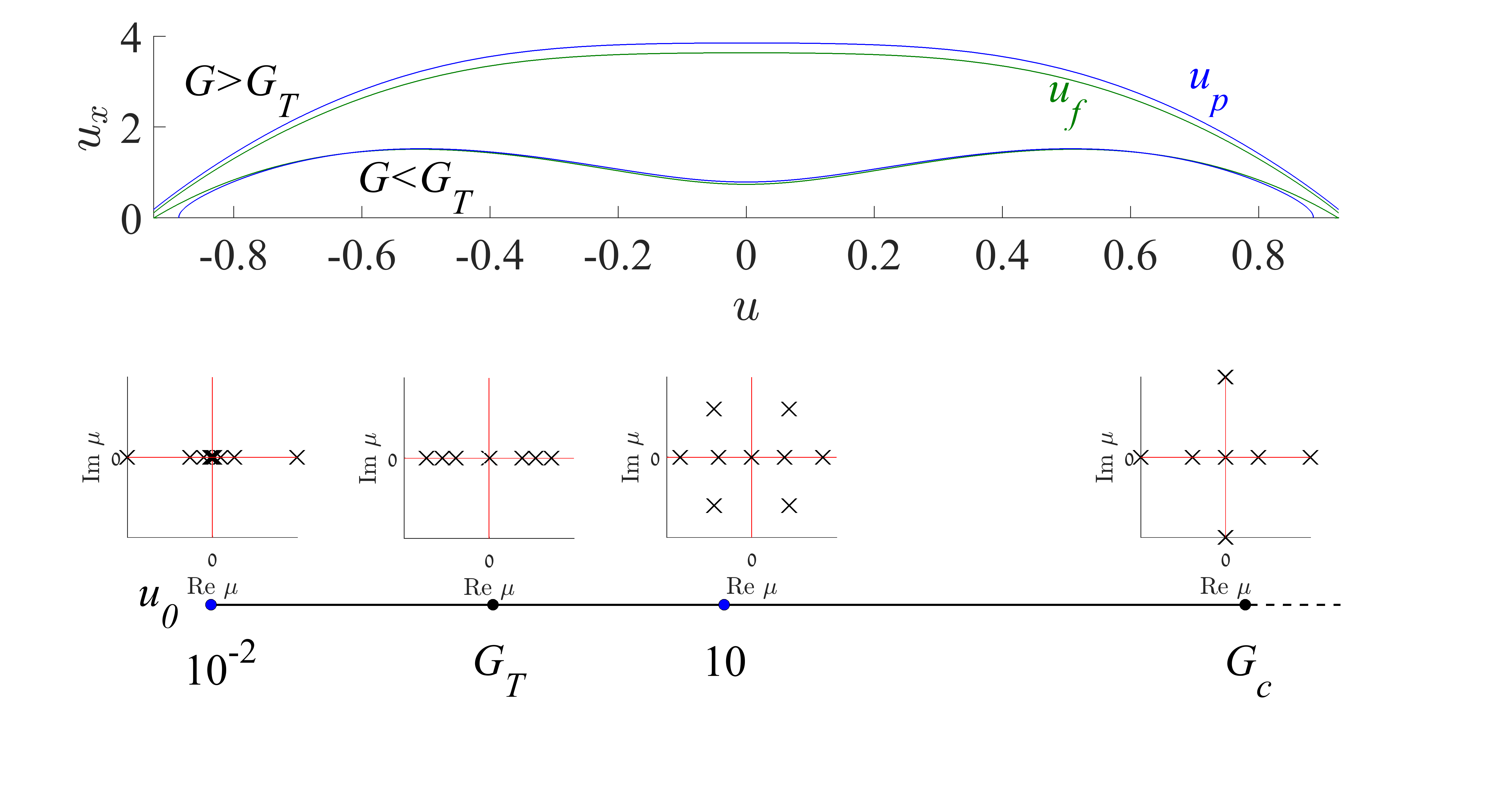}
		\caption{Configuration of the spatial eigenvalues ($\mu$) about the $u=u_0$ state at $G$ values indicated by dots. Top panel shows the respective phase plane for the $u$ field of front ($u_f$) and (one half) periodic ($u_p$) profiles that have been obtained in Fig.~\ref{fig:fig1}(b). Other parameters as in Fig.~\ref{fig:fig1}(b).} \label{fig:fig2}
	\end{figure*}

	Periodic solutions ($u_p$) bifurcate sub-critically from $u=u_0$ at the onset $G=G_c$, that is toward the linearly stable portion of $u_0$, and thus are unstable, as demonstrated in Fig.~\ref{fig:fig1}(b). They gain stability via a saddle-node bifurcation about $G=G_{SN}$ and fold back to large values of $G$. The profiles about the intermediate phase of this stable periodic solutions coincide with those of the coexisting front solutions ($u_f$) that are superimposed in Fig.~\ref{fig:fig1}(b). This similarity implies a generic property that we recover using spatial dynamics analysis, i.e., omitting temporal derivatives, rewriting~\eqref{eq:rmodel} as first order ordinary differential equations, and looking at steady state solutions and envision space as time in a similar fashion as for the Ohta-Kawasaki analysis used for charged polymers~\cite{gavish2017spatially}. Linearizion about $u_0$ reveals twelve  eigenvalues ($\mu$) of which quartic multiplicity is always at zero {(Re $\mu$ = Im $\mu=0$)}, four are always reals, and the remaining four play a crucial role determining the signature of the intermediate phase in profiles that are shown in Fig.~\ref{fig:fig1}(b). At the finite wavenumber instability onset ($G=G_c$), these four eigenvalues lie in a double multiplicity and purely imaginary {(Re $\mu=0$)}~ which is a signature of a reversible Hopf bifurcation (in space)~\cite{champneys1998homoclinic,thiele2013localized,gavish2017spatially}, as shown in Fig.~\ref{fig:fig2}. For $G<G_c$ these eigenvalues split and for $G_T<G<G_c$ persist as complex conjugated, implying the connection to the unstable manifold of unstable limit cycles or a periodic solutions in space that bifurcate from $G=G_c$, as shown in the top panel of Fig.~\ref{fig:fig2}. At $G=G_T$, where
	\begin{equation}\label{gT}
	G_T=\frac{\epsilon^2\gamma(\tau +1)(\beta\xi-\beta-1)^2}{4(\lambda-\epsilon\zeta+\epsilon\zeta^2+2\zeta\sqrt{\epsilon\lambda})^2},
	\end{equation}
	the eigenvalues collide on the real axis and since all eigenvalues are real the impact of the limit cycles vanishes, namely for $G<G_T$ the attractor about $u_0$ is hyperbolic in space and thus orbit spends more ``time'' near $u_0$ making the inflection in the profile visible, as is also shown in the top panel of Fig.~\ref{fig:fig2}. The analytical result for $G_T$ (dashed line in Fig.~\ref{fig:fig1}(a)) agrees well with the numerical computations (dotted line). However, the situation is not necessarily generic and under different parameter setting ($G_T \ll G_c$), the spatial structure about the $u=0$ state can be expected to show also decayed oscillations in space~\cite{yochelis2006reciprocal}, yet, this is beyond the scope here and will be addressed elsewhere. Notably, although mathematically it is obvious that $G_T<G_c$ due to temporal linear stability, the result is also consistent physically: condition $G_T>G_c$ implies reference energies are lower than the thermal energy ($\zeta<k_BT$) which is unrealistic inequality, see SM for details and definitions.
	
	\section{Implications to OPV}
	
	To relate the above analysis in one space to the role of $u_0$ in certain aspects of performance of OPVs, we consider two dimensional domain $[x\times y]\in [L_x \times L_y]$ and supplement Eqs.~\eqref{eq:rmodel} with periodic boundary conditions in $x$ direction, and in $y$ direction of charge flux ($p$ and $n$) and potential~\cite{scott1999charge}:
	
	\begin{equation}\label{eq:bc}
	\nonumber	\left(J^u_y,J^\chi_y,J^p_y,J^n_y,\phi \right)\big\vert_{y=0}=\left(0,0,-D_p p\frac{\partial \phi}{\partial y},0,\frac{V}{2}\right)
	\end{equation}
	and
	\begin{equation}\label{eq:bc}
	\nonumber	\left(J^u_y,J^\chi_y,J^p_y,J^n_y,\phi \right)\big\vert_{y=L_y}=\left(0,0,0,-D_n n\frac{\partial \phi}{\partial y},-\frac{V}{2}\right),
	\end{equation} 
	where $J_y$ for each field is given in the SI, $V$ is the applied voltage between the electrodes which here we take to be under short circuit conditions, i.e, $V=0$.  
	
\begin{figure*}[tp]
		(a)\includegraphics[width=0.155\textwidth]{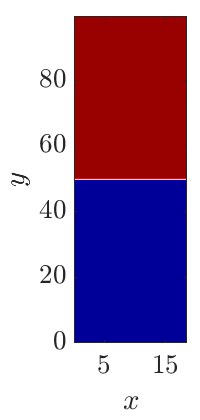} 
		(b)\includegraphics[width=0.41\textwidth]{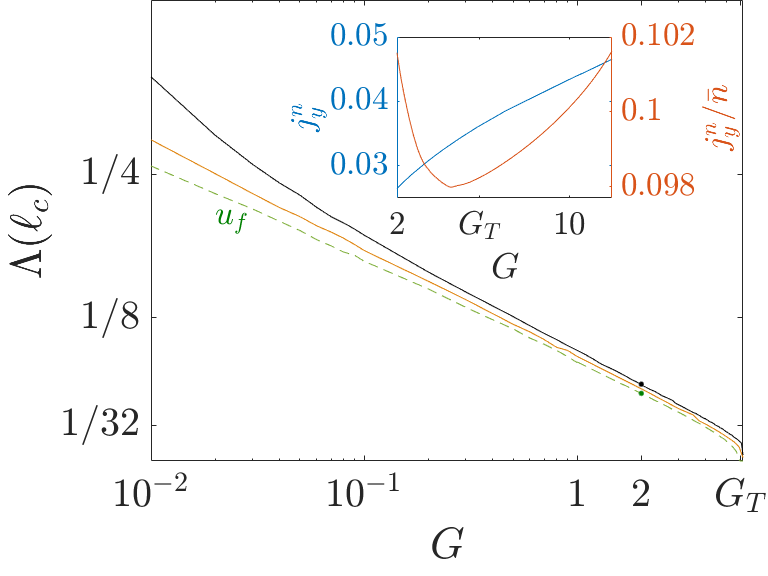}
		(c)\includegraphics[width=0.205\textwidth]{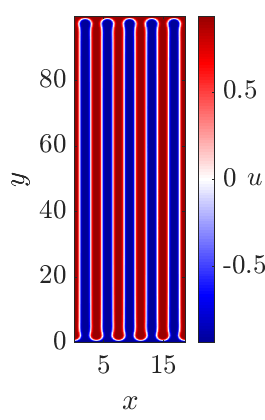}
		\caption{(a) Bilayer steady state solution obtained for $G=2$ and $V=0$, where $L_x=5\ell_c$ and $\ell_c=2\pi/k_c\approx3.73$. (b) Width of $u_0$ scaled by $\ell_c$ as a function of $G$ for bilayer solutions. The bottom (dashed) line represents the width with no-flux boundary conditions as in one-space dimension while top two line represent short circuit conditions for $L_y=200$ and $L_y=100$ (top line). The inset shows current density $j^n_y$ through the boundaries (left vertical axis) and a normalized current density {$j^n_y/\bar{n}$ (right vertical axis).}~ Here computations have been performed for $L_y=100$. (c) Stripe steady state solution obtained under as in (a). Other parameter are as in Fig.~\ref{fig:fig1}.} \label{fig:fig3} 
	\end{figure*}

	First, we examine the effect of applied voltage on a bilayer geometry, i.e, solution $u_f$ (Fig.~\ref{fig:fig3}(a)). Change in the boundary condition for $u_f$ results in current of charges (finite size effect) which in turn, increases the width of $u_0$ and that approaches $\Lambda$ that is obtained in one space dimension when $L_y \to \infty$, as demonstrated in Fig.~\ref{fig:fig3}(b). The widening of $\Lambda$ is profound for $G \ll G_T$ and decreases as $G \to G_T$. In fact, this widening of $u_0$ becomes crucial for a spatially extended periodic solutions in $y$ direction, i.e., ideal (stripe) morphology, as shown in Fig.~\ref{fig:fig3}(c); stripe morphology is obtained by extending the periodic solutions ($u_p$) in $y$ and to keep fidelity to OPV BHJ introduce an additional strip of the phases near electrodes. 
	
\begin{figure}[tp]
		\includegraphics[width=0.45\textwidth]{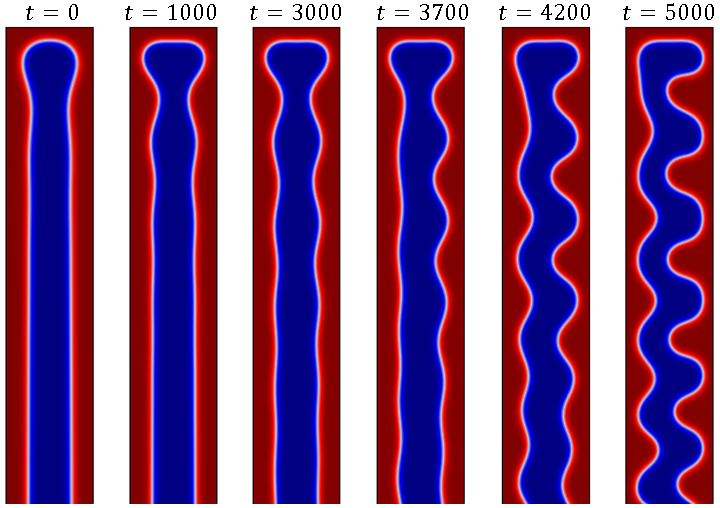}
		\caption{Simulation at $G=20$, showing stripe evolution due to zig-zag instability over one-half vertical direction ($y\in [L_y/2,L_y]$, where $L_y=100$) and one period ($x\in [0,\ell_c]$) in lateral direction. Parameters and boundary conditions as in Fig.~\ref{fig:fig3}.} \label{fig:fig4} 
	\end{figure}

	The increase in $\Lambda$ results in an instability to stripes so that stripes involving $u_0$ persist only near $G_T$, here from about $G\approx2$ and up to $G_T$, while for $G>G_T$ the intermediate phase, $u_0$, vanishes as in the one dimensional case. In particular, in this regime the quantitative analogy between the bilayer and stripes in respect to $\Lambda$ also holds here, thus the latter can be efficiently deduced from analysis in one-space dimension, either from $u_f$ or $u_p$. As expected, the trend of current density, {$j^n_y=L_x^{-1}\int J^n_y{\rm d}x$, at short circuit}~ conditions increases with $G$ (see inset in Fig.~\ref{fig:fig3}(b)) while the ratio {$j_y^n/\bar{n}$, where $\bar{n}=L_x^{-1}L_y^{-1}\int n(x,y) {\rm d}x {\rm d}y$}, that indicates outflow vs. creation (due to $G$) of charge is essentially constant. In fact, this region about $G_T$ of constant normalized current indicates stability of periodic solutions (ideal BHJ), a direct consequence of tri-stable system. Specifically, outside this region at higher $G$ values, periodic stripes that are not influenced by the attractor of the $u_0$ become unstable to zig-zag~\cite{gavish2017spatially} leading to deformed morphology, as shown in Fig.~\ref{fig:fig4}. Notably, a detailed analysis of stripe instability is out of the scope here and will be discussed in detail elsewhere.

	\section{Discussion}
	
	We have formulated a distinct mean-field model that enables to capture binary (bi-stable) and ternary (tri-stable) bulk heterojuction (BHJ). We showed that although the width of the intermediate phase ($\Lambda$) in BHJ depends on illumination strength ($G$) and applied voltage ($V$) its trend can be efficiently deduced from a relatively simply analysis in one space dimension as long as $L_y\gg \Lambda$, which is consistent with device-wise applications. Moreover, we conjecture that existence of the intermediate phase impacts stability of BHJ striped morphology -- a criterion that is valuable to design efficient and stable BHJ based organic photovoltaic devices. In a broader context of material science, our framework opens new vistas to a wider range of electrically charged amphiphilic-type mixtures to tackle their stability and evolution, example of which include emulsions, polyelectrolytes, and ionic liquids.
	
\begin{acknowledgements}
	We thank Profs. Iris Visoly-Fisher and Rafi Shikler (Ben-Gurion University of the Negev) for stimulating discussions on OPV and to Prof. Hannes Uecker (Universit\"{a}t Oldenburg) for help in implementation of pde2path package. The research was done in the framework of the Grand Technion Energy Program (GTEP) and of the BGU Energy Initiative Program, and supported by the Adelis Foundation for renewable energy research.
\end{acknowledgements}

%

\end{document}